\documentclass[article,amsmath,amssymb,nofootinbib,aps,prc,superscriptaddress]{revtex4-2}

\usepackage{graphicx}
\usepackage{epstopdf}
\usepackage{caption}
\usepackage{color}
\usepackage{bm}
\usepackage{textcomp} 
\usepackage{subfigure}
\usepackage{ulem}
\usepackage{color}
\usepackage{mathtools}
\usepackage{amsmath}
\graphicspath{{figures/}}
\usepackage{diagbox}
\usepackage{slashbox}
\usepackage{booktabs}
\usepackage{multirow}

\usepackage{array}
\usepackage{amssymb}

\begin{document}

\title{Observation of the electromagnetic field effect via charge-dependent directed flow in heavy-ion collisions at the Relativistic Heavy Ion Collider}

\affiliation{Abilene Christian University, Abilene, Texas   79699}
\affiliation{Alikhanov Institute for Theoretical and Experimental Physics NRC "Kurchatov Institute", Moscow 117218}
\affiliation{Argonne National Laboratory, Argonne, Illinois 60439}
\affiliation{American University in Cairo, New Cairo 11835, Egypt}
\affiliation{Ball State University, Muncie, Indiana, 47306}
\affiliation{Brookhaven National Laboratory, Upton, New York 11973}
\affiliation{University of Calabria \& INFN-Cosenza, Rende 87036, Italy}
\affiliation{University of California, Berkeley, California 94720}
\affiliation{University of California, Davis, California 95616}
\affiliation{University of California, Los Angeles, California 90095}
\affiliation{University of California, Riverside, California 92521}
\affiliation{Central China Normal University, Wuhan, Hubei 430079 }
\affiliation{University of Illinois at Chicago, Chicago, Illinois 60607}
\affiliation{Creighton University, Omaha, Nebraska 68178}
\affiliation{Czech Technical University in Prague, FNSPE, Prague 115 19, Czech Republic}
\affiliation{National Institute of Technology Durgapur, Durgapur - 713209, India}
\affiliation{ELTE E\"otv\"os Lor\'and University, Budapest, Hungary H-1117}
\affiliation{Frankfurt Institute for Advanced Studies FIAS, Frankfurt 60438, Germany}
\affiliation{Fudan University, Shanghai, 200433 }
\affiliation{Guangxi Normal University, Guilin, China}
\affiliation{University of Heidelberg, Heidelberg 69120, Germany }
\affiliation{University of Houston, Houston, Texas 77204}
\affiliation{Huzhou University, Huzhou, Zhejiang  313000}
\affiliation{Indian Institute of Science Education and Research (IISER), Berhampur 760010 , India}
\affiliation{Indian Institute of Science Education and Research (IISER) Tirupati, Tirupati 517507, India}
\affiliation{Indian Institute Technology, Patna, Bihar 801106, India}
\affiliation{Indiana University, Bloomington, Indiana 47408}
\affiliation{Institute of Modern Physics, Chinese Academy of Sciences, Lanzhou, Gansu 730000 }
\affiliation{University of Jammu, Jammu 180001, India}
\affiliation{Joint Institute for Nuclear Research, Dubna 141 980}
\affiliation{Kent State University, Kent, Ohio 44242}
\affiliation{University of Kentucky, Lexington, Kentucky 40506-0055}
\affiliation{Lawrence Berkeley National Laboratory, Berkeley, California 94720}
\affiliation{Lehigh University, Bethlehem, Pennsylvania 18015}
\affiliation{Max-Planck-Institut f\"ur Physik, Munich 80805, Germany}
\affiliation{Michigan State University, East Lansing, Michigan 48824}
\affiliation{National Research Nuclear University MEPhI, Moscow 115409}
\affiliation{National Institute of Science Education and Research, HBNI, Jatni 752050, India}
\affiliation{National Cheng Kung University, Tainan 70101 }
\affiliation{The Ohio State University, Columbus, Ohio 43210}
\affiliation{Panjab University, Chandigarh 160014, India}
\affiliation{NRC "Kurchatov Institute", Institute of High Energy Physics, Protvino 142281}
\affiliation{Purdue University, West Lafayette, Indiana 47907}
\affiliation{Rice University, Houston, Texas 77251}
\affiliation{Rutgers University, Piscataway, New Jersey 08854}
\affiliation{University of Science and Technology of China, Hefei, Anhui 230026}
\affiliation{South China Normal University, Guangzhou, Guangdong 510631}
\affiliation{Sejong University, Seoul, 05006, South Korea}
\affiliation{Shandong University, Qingdao, Shandong 266237}
\affiliation{Shanghai Institute of Applied Physics, Chinese Academy of Sciences, Shanghai 201800}
\affiliation{Southern Connecticut State University, New Haven, Connecticut 06515}
\affiliation{State University of New York, Stony Brook, New York 11794}
\affiliation{Instituto de Alta Investigaci\'on, Universidad de Tarapac\'a, Arica 1000000, Chile}
\affiliation{Temple University, Philadelphia, Pennsylvania 19122}
\affiliation{Texas A\&M University, College Station, Texas 77843}
\affiliation{University of Texas, Austin, Texas 78712}
\affiliation{Tsinghua University, Beijing 100084}
\affiliation{University of Tsukuba, Tsukuba, Ibaraki 305-8571, Japan}
\affiliation{University of Chinese Academy of Sciences, Beijing, 101408}
\affiliation{Valparaiso University, Valparaiso, Indiana 46383}
\affiliation{Variable Energy Cyclotron Centre, Kolkata 700064, India}
\affiliation{Warsaw University of Technology, Warsaw 00-661, Poland}
\affiliation{Wayne State University, Detroit, Michigan 48201}
\affiliation{Yale University, New Haven, Connecticut 06520}

\author{M.~I.~Abdulhamid}\affiliation{American University in Cairo, New Cairo 11835, Egypt}
\author{B.~E.~Aboona}\affiliation{Texas A\&M University, College Station, Texas 77843}
\author{J.~Adam}\affiliation{Czech Technical University in Prague, FNSPE, Prague 115 19, Czech Republic}
\author{J.~R.~Adams}\affiliation{The Ohio State University, Columbus, Ohio 43210}
\author{G.~Agakishiev}\affiliation{Joint Institute for Nuclear Research, Dubna 141 980}
\author{I.~Aggarwal}\affiliation{Panjab University, Chandigarh 160014, India}
\author{M.~M.~Aggarwal}\affiliation{Panjab University, Chandigarh 160014, India}
\author{Z.~Ahammed}\affiliation{Variable Energy Cyclotron Centre, Kolkata 700064, India}
\author{A.~Aitbaev}\affiliation{Joint Institute for Nuclear Research, Dubna 141 980}
\author{I.~Alekseev}\affiliation{Alikhanov Institute for Theoretical and Experimental Physics NRC "Kurchatov Institute", Moscow 117218}\affiliation{National Research Nuclear University MEPhI, Moscow 115409}
% ...

\collaboration{STAR Collaboration}\noaffiliation

\begin{abstract}
The deconfined quark-gluon plasma (QGP) created in relativistic heavy-ion collisions enables the exploration of the fundamental properties of matter under extreme conditions. Non-central collisions can produce strong magnetic fields on the order of $10^{18}$ Gauss, which offers a probe into the electrical conductivity of the QGP.
In particular,  quarks and anti-quarks carry opposite charges and receive contrary electromagnetic forces that alter their momenta. This phenomenon can be manifested in the collective motion of final-state particles, specifically in the rapidity-odd directed flow, denoted as $v_1(\mathsf{y})$.
Here we present the charge-dependent measurements of $dv_1/d\mathsf{y}$ near midrapidities for  $\pi^{\pm}$, $K^{\pm}$, and $p(\bar{p})$ in Au+Au and isobar ($_{44}^{96}$Ru+$_{44}^{96}$Ru and  $_{40}^{96}$Zr+$_{40}^{96}$Zr) collisions at $\sqrt{s_{\rm NN}}=$
200 GeV, and in Au+Au collisions at 27 GeV, recorded by the
STAR detector at the Relativistic Heavy Ion Collider. 
The combined dependence of the $v_1$ signal on collision system, particle species, and collision centrality can be qualitatively and semi-quantitatively understood as several effects on constituent quarks. While the results in central events can be explained by the $u$ and $d$ quarks transported from initial-state nuclei, those in peripheral events reveal the impacts of the electromagnetic field on the QGP. Our data put valuable constraints on the electrical conductivity of the QGP in theoretical calculations.
\end{abstract}
    
\maketitle

\section{Introduction}
By colliding two heavy nuclei at high center-of-mass energies ($\sqrt{s_{\rm NN}}$), experiments at the BNL Relativistic Heavy Ion Collider (RHIC) and the CERN Large Hadron Collider (LHC) can create a  medium of liberated quarks and gluons, a state of matter known as the quark-gluon plasma (QGP)~\cite{RevModPhys.89.035001}. The QGP dominated the early Universe about a microsecond after the Big Bang~\cite{BigBang-Kolb}, and its recreation in the laboratory provides a unique opportunity to study the fundamental properties of matter under extreme conditions. 
In these ultra-relativistic collisions, the nuclear fragments pass by each other, generating very strong magnetic fields, on the order of $10^{18}$ Gauss~\cite{Voronyuk:2011jd,Deng:2012pc,Zhao:2019crj,Kharzeev:2007jp,MCLERRAN2014184,Wang:2021oqq,Grayson:2022asf}, the evolution of which, in the presence of a QGP, must be described in conjunction with the QGP's electromagnetic properties. The presence of a strong magnetic field also facilitates the study of some novel phenomena related to the restoration of fundamental symmetries of quantum chromodynamics (QCD)~\cite{Chernodub:2010qx,Frolov:2010wn,Chernodub:2011mc,Burnier:2011bf,Huang:2015oca,Kharzeev:2007jp,Hattori:2016emy}. For example, the chiral magnetic effect (CME) predicts a charge separation along the direction of magnetic field due to the chirality imbalance and chiral symmetry restoration in the QGP~\cite{Kharzeev:2020jxw,Fukushima:2008xe,Kharzeev:2015znc}. If confirmed, the CME in heavy-ion collisions will uncover the local parity and charge-parity violation in the strong interaction~\cite{Kharzeev:2007jp}. 
The strong magnetic field could also interact with the QCD matter in other ways, such as, providing a catalyst for chiral symmetry breaking~\cite{Gusynin:1994re}, causing the synchrotron radiation from quarks~\cite{Tuchin:2010vs},  differentiating the chiral and deconfinement phase transitions in the QCD phase diagram~\cite{Mizher:2010zb}, and modifying the collectivity of a QGP~\cite{Mohapatra:2011ku,Tuchin:2011jw,Das:2016cwd,Dubla:2020bdz,Gursoy:2018yai,Gursoy:2014aka,Nakamura:2022ssn}.   

\begin{figure}[htbp]
\vspace*{-0.1in}
\includegraphics[scale=0.4]{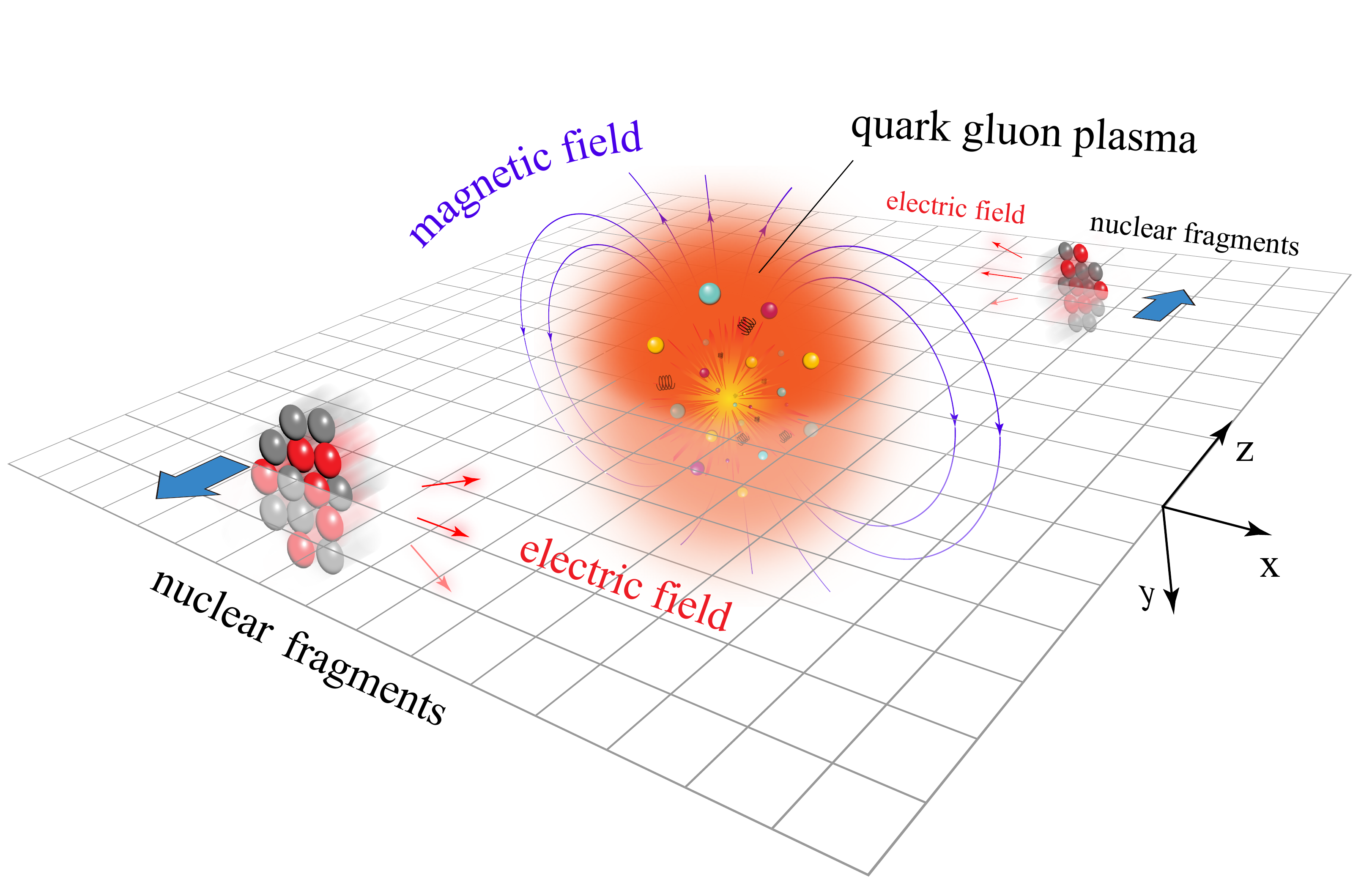}
\captionof{figure}{Sketch of a heavy-ion collision in the lab frame.
The impact parameter and the beam direction are along the $x$ and $z$ axes, respectively. 
The $x$-$z$ plane is called the reaction plane. Participating nucleons in the overlap region create a hot and dense medium of  quark-gluon plasma.
Spectator nuclear fragments generate 
strong electromagnetic fields. }
	\label{fig:illustration}
\end{figure}

\begin{figure}[htbp]
\vspace*{-0.1in}
\includegraphics[scale=0.4]{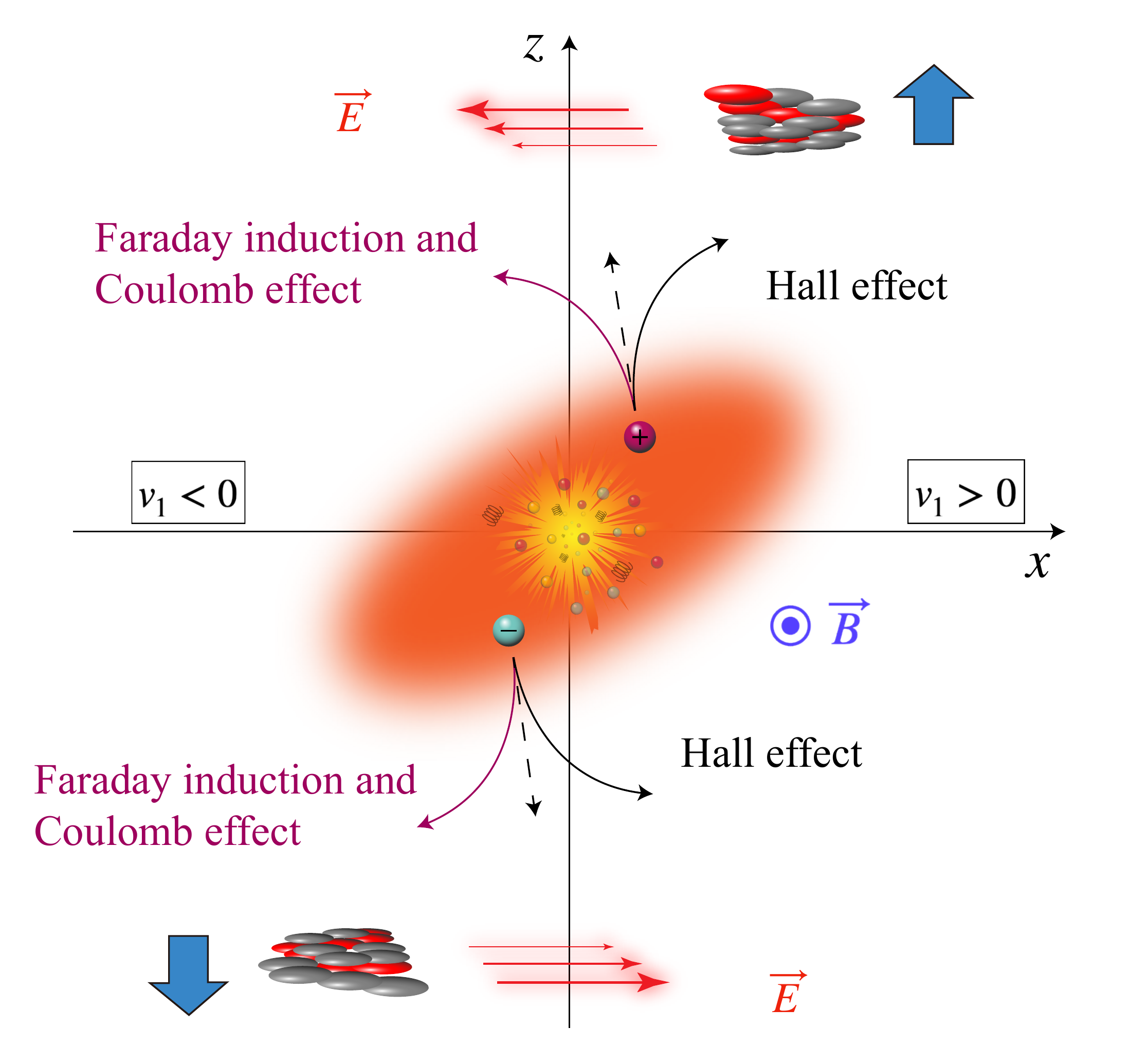}
\captionof{figure}{Schematic overhead view (along the $y$ axis) of a heavy-ion collision in the lab frame. The dashed lines represent the motion of quarks due to the QGP expansion.
The black (purplish red) curved lines suggest the paths that quarks would follow due to the Lorentz force (Coulomb and Faraday field) alone.}
	\label{fig:illustration2}
\end{figure}

Direct evidence of the electromagnetic field in the QGP is elusive
because the magnetic field magnitude drops rapidly with time in the vacuum until the QCD medium is formed, after which the field couples with the induced electric current in the QGP.
Previously, the  Coulomb effect in asymmetric Cu+Au collisions was  observed via charge-dependent rapidity-even directed flow, $v_1^{\rm even}(\mathsf{y})$~\cite{STAR:2016cio,STAR:2017ykf}.
Directed flow ($v_1$) is defined as the first Fourier coefficient of the particle azimuthal distribution relative to the reaction plane (the $x$-$z$ plane in Fig.~\ref{fig:illustration})~\cite{Poskanzer:1998yz,Voloshin:2016ppr}, 
\begin{equation}
E\frac{d^3N}{d^3p} = \frac{1}{2\pi}\frac{d^2N}{p_Tdp_Td\mathsf{y}} \left( 1 + \sum_{n=1}^{\infty} 2v_n\cos n(\phi - \Psi) \right),    
\end{equation}
where $p_T=\sqrt{p_x^2+p_y^2}$ is transverse momentum, while $\phi$ and $\Psi$ are the azimuthal angles of a particle and the reaction plane, respectively. 
Note that rapidity ($\mathsf{y}$) and $p_z$ bear the same sign. $v_1(\mathsf{y})$ can be uniquely expressed as a combination of two components: an even function of $\mathsf{y}$ and an odd function.
Recent studies suggest that the charge-dependent $v_1^{\rm odd}(\mathsf{y})$ can serve as a probe to the electromagnetic field in symmetric heavy-ion collisions~\cite{Gursoy:2014aka,Gursoy:2018yai,Das:2016cwd,Nakamura:2022ssn}; we explore this approach below. Hereafter, $v_1(\mathsf{y})$ implicitly refers to the odd component, which comes from the initial tilt of the QGP (Fig.~\ref{fig:illustration2}) and is sensitive to the Equation of State~\cite{Luzum:2010fb,Heinz:2013th}. Measurements of $v_1$ have been extensively performed over past decades at both RHIC and the LHC experiments~\cite{STAR:2005btp,STAR:2008jgm,STAR:2011gzz,STAR:2011hyh,STAR:2014clz,STAR:2016cio,STAR:2017okv,STAR:2019clv,STAR:2020hya,ALICE:2013xri}.
It is common practice to present $dv_1/d\mathsf{y}$ because of the linear $\mathsf{y}$ dependence of $v_1$ near midrapidities in those experiments. 

Figure \ref{fig:illustration2} illustrates the overhead view of a heavy-ion collision, where the 
longitudinal expansion of the QGP has the same effect on $v_1$ for quarks with opposite charges. This degeneracy will be lifted by electromagnetic effects. For a positively charged quark, the Lorentz force in the Hall effect (black solid lines in Fig.~\ref{fig:illustration2}) increases its $v_1$ at $\mathsf{y}>0$, and decreases its $v_1$ at $\mathsf{y}<0$, namely increasing its $dv_1/d\mathsf{y}$. For a negatively charged quark, the Hall effect does the opposite, and decreases its $dv_1/d\mathsf{y}$, contributing a positive $\Delta(dv_1/d\mathsf{y})$~\footnote{In our notation, $\Delta(dv_1/d\mathsf{y})$ is the $dv_1/d\mathsf{y}$ difference between positively charged particles and their negatively charged anti-particles. For example, proton $\Delta(dv_1/d\mathsf{y})$ means the $dv_1/d\mathsf{y}$ difference between protons and anti-protons.}.
In addition, the fast decay of the magnetic field in the QGP will induce an electric field, known as the Faraday induction effect,
which, together with the electric field from spectator protons, renders a negative $\Delta(dv_1/d\mathsf{y})$ for quarks (purplish red lines). 
Theoretical calculations predict that the Faraday induction + Coulomb effect will dominate over the Hall effect on light quarks~\cite{Gursoy:2014aka,Gursoy:2018yai,Nakamura:2022ssn}, and their roles are reversed on charm quarks that have relatively early formation time and long relaxation time~\cite{Das:2016cwd}. The net effect for quarks will be translated into a finite $\Delta(dv_1/d\mathsf{y})$ between positively and negatively charged hadrons through quark coalescence. In the coalescence
picture, the $v_n$ of
the resulting mesons or baryons is roughly equal to the summed $v_n$ of their constituent quarks~\cite{Molnar:2003ff,STAR:2017okv,Dunlop:2011cf,Goudarzi:2020eoh}. 

As a well-recognized paradigm,  constituent quarks refer to valence quarks enveloped by a cloud of sea-quarks and gluons,  effectively gaining an enhanced mass.
Mounting evidence  supports the coalescence of two or three constituent quarks within the dense QGP medium, forming mesons or baryons.
Although not understood at a fundamental level, this mechanism has demonstrated remarkable success in elucidating  the multiplicity dependence of yields, spectra, and collective motions in heavy-ion collisions~\cite{Molnar:2003ff,STAR:2017okv,Dunlop:2011cf,Goudarzi:2020eoh,PhysRevC.67.064902,PhysRevLett.90.202303,PhysRevLett.90.202302}.
In this paper, we differentiate between two sources of constituent quarks: those generated as $q$$\bar q$ pairs and those transported from
the initial-state nuclei to midrapidities.
Transported quarks convey information from the incident nucleons and undergo the entire system evolution. Conversely, produced quarks are likely to form at various stages.
In the high-$\sqrt{s_{\rm NN}}$ limit, most $u$ and $d$ quarks are produced, whereas in the low-$\sqrt{s_{\rm NN}}$ limit, most of them are presumably transported. The fraction of transported $u$($d$) in all $u$($d$) quarks can be estimated, e.g., following Boltzmann statistics with the experimentally measured temperature and baryon chemical potential of the collision system~\cite{Goudarzi:2020eoh}.

\begin{figure}[htbp]
\vspace*{-0.1in}
{\includegraphics[scale=0.6]{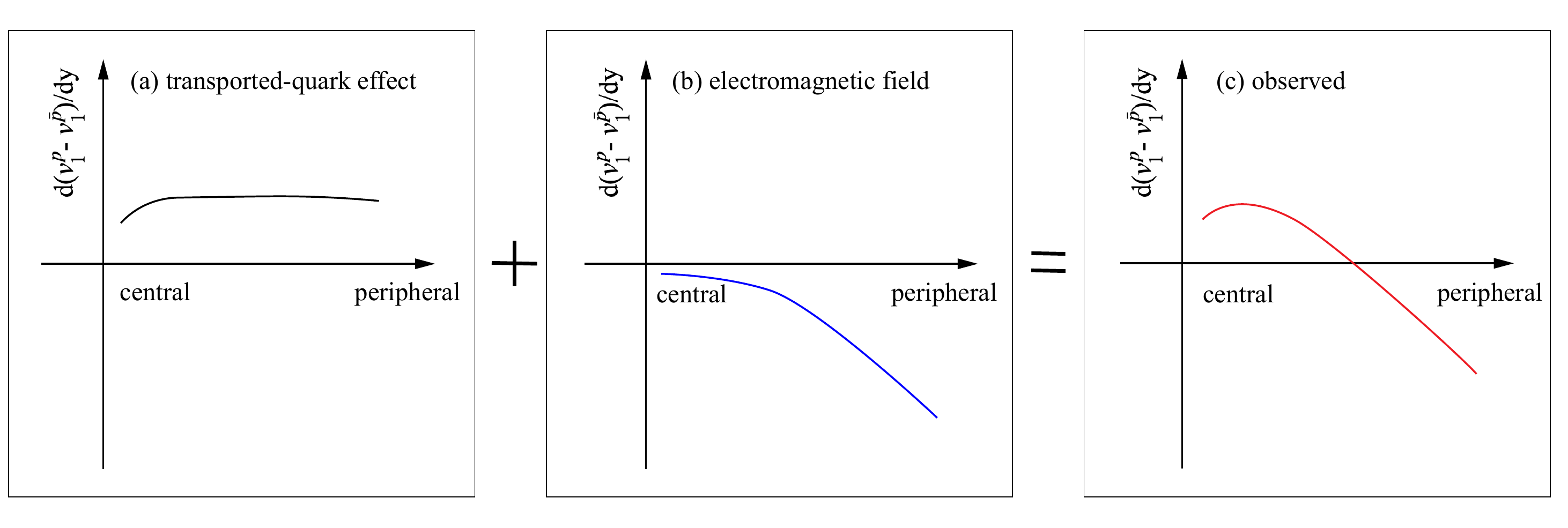}}   	
\captionof{figure}{Illustration of different contributions to the proton $\Delta(dv_1/dy)$  versus centrality.  Panel (a) depicts the transported-quark effect. Panel (b) sketches the electromagnetic field contribution, dominated by the Faraday induction + Coulomb effect. Panel (c)  speculates the superposition of the two effects in the final observable.}
\label{fig:Cartoon4Discussion}
\end{figure}

Experimental efforts have been made to search for the electromagnetic field effect in symmetric collisions, such as the $dv_1/d\mathsf{y}$ measurements for $D^0$ and $\bar{D}^0$ in Au+Au collisions at $\sqrt{s_{\rm NN}}= 200$ GeV by the STAR experiment~\cite{STAR:2019clv}. In addition, the ALICE Collaboration has conducted similar measurements for charged hadrons and $D^0$ mesons in Pb+Pb collisions at $\sqrt{s_{\rm NN}}= 5.02$ TeV~\cite{ALICE:2019sgg}. While the $\Delta(dv_1/d\mathsf{y})$ results for $D^0$ mesons are limited in significance by lack of statistics, those for charged hadrons (consisting mostly of light quarks) in 10-40\% centrality show positive values~\cite{STAR:2019clv,ALICE:2019sgg}, which contradicts the theoretical expectation for the Faraday induction + Coulomb effect. 
Positive $\Delta(dv_1/d\mathsf{y})$ values between protons and anti-protons have also been obtained in semi-central Au+Au collisions at several RHIC  energies~\cite{STAR:2014clz,STAR:2017okv}, and are attributed to  transported quarks. The transported $u$ and $d$ quarks acquire different azimuthal anisotropy than the pair-produced quarks in later stages, and they contribute to protons ($uud$) but not anti-protons ($\bar{u}\bar{u}\bar{d}$).  Calculations from UrQMD~\cite{Guo:2012qi}, AMPT~\cite{Nayak:2019vtn}, and a hydrodynamic model with an expanding fireball of inhomogeneous baryons~\cite{Bozek:2022svy} indicate that transported quarks have positive  $dv_1/d\mathsf{y}$, and hence should give a positive contribution to $\Delta(dv_1/d\mathsf{y})$ between protons and anti-protons, as demonstrated in Fig.~\ref{fig:Cartoon4Discussion}(a).
A similar effect in the elliptic flow ($v_2$) difference between protons and anti-protons has also been observed in the RHIC Beam Energy Scan (BES) data~\cite{Dunlop:2011cf,Goudarzi:2020eoh}.  
On the other hand, Fig.~\ref{fig:Cartoon4Discussion}(b) shows the qualitative trend of the electromagnetic field effect on the proton $\Delta(dv_1/d\mathsf{y})$, in view of the more spectator protons in more peripheral collisions. If the Faraday induction + Coulomb effect dominates over the Hall effect and the transported-quark effect in peripheral collisions, a sign change of the proton $\Delta(dv_1/d\mathsf{y})$ could occur from positive in central events to negative in peripheral ones, as illustrated in panel (c).
Therefore, the negative $\Delta(dv_1/d\mathsf{y})$ between protons and anti-protons can serve as a signature of the Faraday induction + Coulomb effect.

Similar to $p$ and $\bar{p}$ in Fig.~\ref{fig:Cartoon4Discussion}, the $\Delta(dv_1/d\mathsf{y})$ between $K^+$ and $K^-$ could also change sign as a function of centrality,
since transported $u$ quarks increase the $dv_1/d\mathsf{y}$ for only $K^+(u\bar{s})$ but not $K^-(\bar{u}s)$,
giving a positive contribution to the kaon $\Delta(dv_1/d\mathsf{y})$, while the electromagnetic field plays an opposite role that grows stronger in more peripheral events.
Both $\pi^+(u\bar{d})$ and $\pi^-(\bar{u}d)$
are affected by transported quarks.
Since the gold ion ($^{197}_{79}\mathrm{Au}$) is neutron($udd$) rich, there are more $d$ quarks than $u$ quarks transported in Au+Au collisions, and thus $\pi^-$ is more influenced than $\pi^+$~\cite{Dunlop:2011cf},
leading to a negative contribution to the $\Delta (dv_1/d\mathsf{y})$  between $\pi^+$ and $\pi^-$. As a side note, the $v_2$ difference  between $\pi^+$ and $\pi^-$ in the BES data has been quantitatively explained by the transported-quark effect~\cite{Goudarzi:2020eoh,PhysRevC.93.014907}. 
Thus, the transported-quark effect and the Faraday induction + Coulomb effect work in the same direction for the pion $\Delta (dv_1/d\mathsf{y})$, and cannot be distinguished. 
In this work, we report the $dv_1/d\mathsf{y}$ measurements  for  $\pi^{\pm}$, $K^{\pm}$, and $p(\bar{p})$ in Au+Au collisions at $\sqrt{s_{\rm NN}}=$ 27 and 200 GeV, and in isobar ($_{44}^{96}$Ru+$_{44}^{96}$Ru and  $_{40}^{96}$Zr+$_{40}^{96}$Zr) collisions at 200 GeV, with the expectation that
the large data sets may reveal a sign change in the proton and kaon $\Delta (dv_1/d\mathsf{y})$ as a function of centrality.

\section{Experiment and methodology}

The STAR experiment collected large data samples of
minimum-bias-trigger events of Au+Au 
collisions at $\sqrt{s_{\rm NN}}=$ 200 GeV in
2014 and 2016,
and of isobar collisions at $\sqrt{s_{\rm NN}}=$ 200 GeV and Au+Au collisions at $\sqrt{s_{\rm NN}}=$ 27 GeV in 2018. The Time
Projection Chamber (TPC)~\cite{ANDERSON2003659}
was used for charged particle tracking within pseudorapidity $|\eta|<1$, with full $2\pi$ azimuthal coverage. 
For each event, 
the primary vertex position along the beam direction (the $z$ axis) is reconstructed with both the TPC ($V_{z,{\rm TPC}}$) and the Vertex Position Detectors ($V_{z,{\rm VPD}}$)~\cite{LLOPE201423}. The radial distance between the primary vertex and the $z$ axis ($V_r$) is evaluated with the TPC.
To ensure the event quality, each event is required to have a vertex position within $|V_{z,{\rm TPC}}|< 30$ cm ($<70$ cm),   $V_r < 2$ cm and $|V_{z,{\rm TPC}}-V_{z,{\rm VPD}}| < 3$ cm ($<4$ cm) for Au+Au collisions at 200 GeV (27 GeV), 
and within $-35 < V_{z,{\rm TPC}} < 25$ cm, $V_r < 2$ cm and $|V_{z,{\rm TPC}}-V_{z,{\rm VPD}}| < 5$ cm for isobar collisions at  200 GeV. The asymmetric $V_{z,{\rm TPC}}$ cuts come from a negative mean value of $\langle V_{z,{\rm TPC}} \rangle = -5$ cm in isobar collisions.  After the vertex selection, we have about 2.2 billion Au+Au events at 200 GeV, 400 million Au+Au events at  27 GeV, 1.7 billion Ru+Ru events and 1.8 billion Zr+Zr events at  200 GeV. Centrality is defined by matching the distribution of the number of charged particles (detected by the TPC within $|\eta|<0.5$) and the one obtained from MC Glauber simulations~\cite{Miller:2007ri,STAR:2008med,STAR:2021mii}. We focus on the centrality range of 0--80\%, where 0 refers to head-on collisions, and 80\% represents very peripheral collisions.

\begin{figure}[htbp]
\vspace*{-0.1in}
{\includegraphics[scale=0.3]{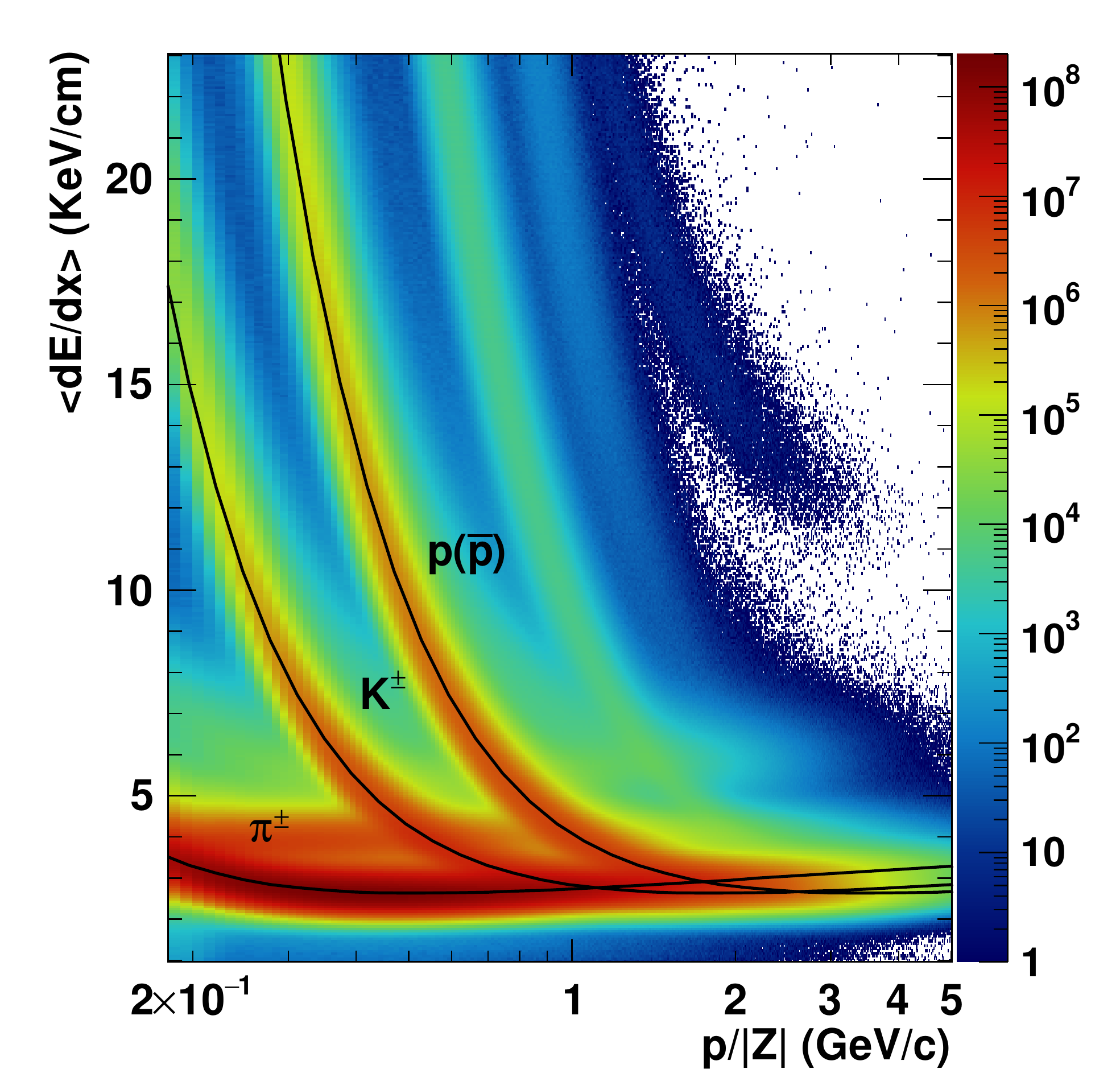}} 
{\includegraphics[scale=0.3]{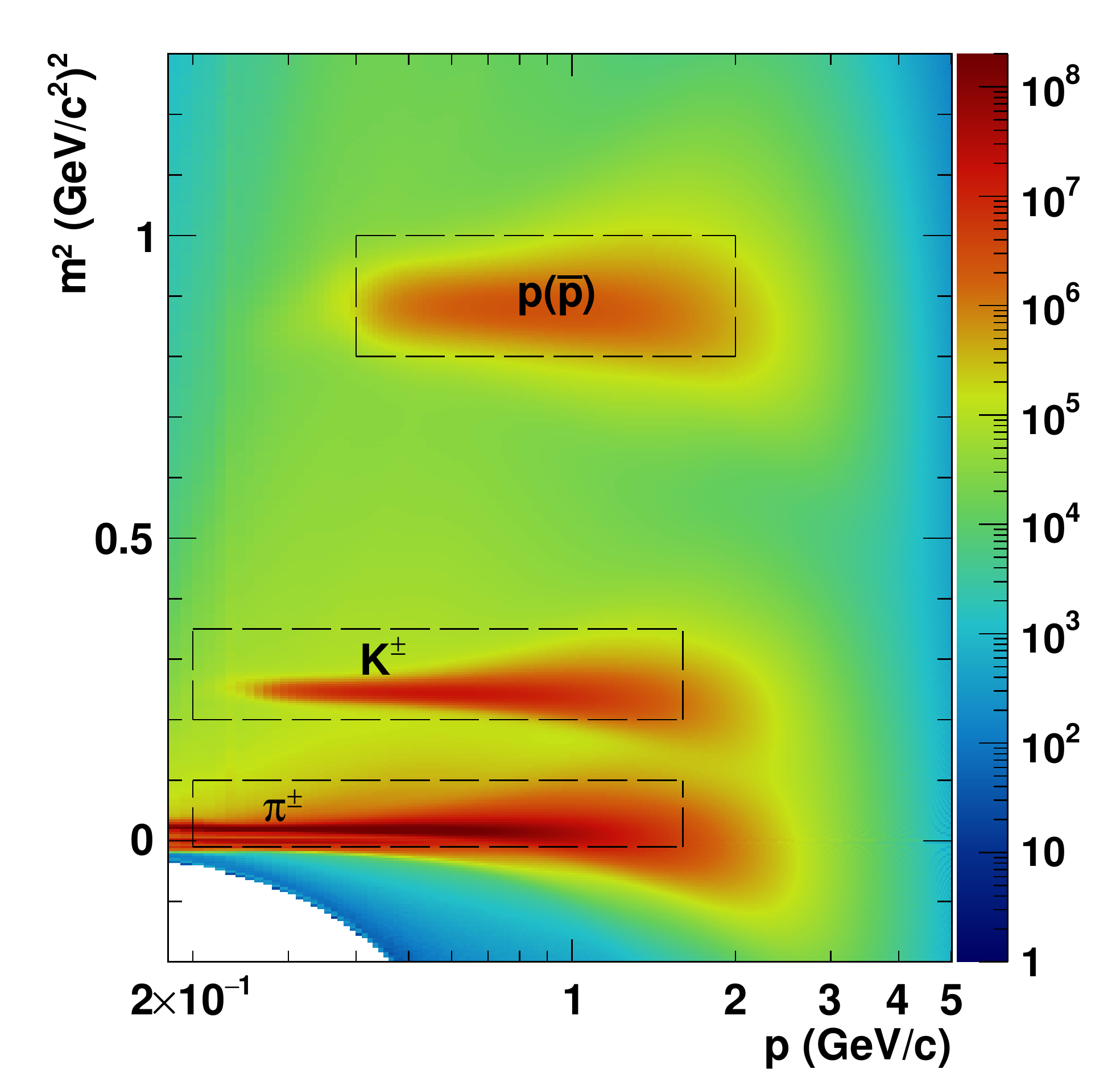}}  	
\captionof{figure}{(Left) Ionization energy loss ($\langle dE/dx \rangle$) of charged particles in the TPC versus magnetic rigidity in Au+Au collisions at $\sqrt{s_{\rm NN}} = 200$ GeV. Solid lines denote the Bichsel functions. (Right)  Mass squared ($m^2$) from the TOF versus momentum. Dashed boxes  indicate the selection criteria  for pions, kaons, and protons.}
\label{fig:TPCdedx}
\end{figure}

In this analysis, tracks within the TPC acceptance are required to have 
at least 15 space points ($N_{\rm hits}$) and
a distance of closest approach (DCA) to the primary vertex less than 2 cm in Au+Au collisions, and  less than 3 cm in isobar collisions. $\pi^\pm$, $K^\pm$, $p$ and $\bar{p}$ are identified based on the truncated mean value of the track energy loss ($\langle dE/dx \rangle$) in the TPC and time-of-flight information from the TOF detector~\cite{LLOPE2004252}. Figure~\ref{fig:TPCdedx} (left) presents an example of $\langle dE/dx \rangle$ versus magnetic rigidity for charged particles in Au+Au collisions at $\sqrt{s_{\rm NN}} = 200$ GeV. Pions, kaons and protons show separate bands below  1 GeV/$c$, and gradually merge at higher momenta. For a specific particle type $i$, the measured $\langle dE/dx \rangle$ can be described by the corresponding Bichsel function~\cite{Bichsel:2006cs}, $\langle dE/dx \rangle_i^{\rm th}$ (solid lines), and we select those candidates with $|n\sigma_i|<2$, where
\begin{equation}
n\sigma_i = \frac{1}{\sigma_R} \ln\bigg(\frac{\left< dE/dx \right>}{\left< dE/dx \right>^{\rm th}_i}\bigg).
\end{equation} 
Here $\sigma_R$ is the momentum dependent $\langle dE/dx \rangle$ resolution. To improve the particle identification at higher momenta, the TOF detector is employed to deduce the mass squared of charged particles. The distribution of mass squared versus momentum in Fig.~\ref{fig:TPCdedx} (right) shows 
the separation between pions, kaons and protons. To ensure the purity of identified particles, the upper bounds of momentum are set to be 2 GeV/$c$  for protons and 1.6 GeV/$c$ for pions and kaons, and the lower bounds of transverse momentum are 0.4 GeV/$c$ for protons and 0.2 GeV/$c$ for pions and kaons.

In practice, $v_1$ is measured with respect to the event plane ($\Psi_{\rm EP}$), an estimated reaction plane, and corrected for its finite resolution ($R\{\Psi_{\rm EP}\}$)~\cite{Poskanzer:1998yz}: 
\begin{equation}
v_1 =  \langle \cos(\phi - \Psi_{\rm EP})  \rangle/R\{\Psi_{\rm EP}\}. \label{Eq: v1FullEP}
\end{equation} 
The average is taken over all particles of interest in an event and then over all events.
In collisions at $\sqrt{s_{\rm NN}}=$ 200 GeV, $\Psi_{\rm EP}$ is determined from the sideward deflection of  spectator neutrons registered by the Zero Degree Calorimeter Shower-Maximum Detectors (ZDC-SMD, $|\eta| > 6.3$)~\cite{ADLER2001488}, and $R\{\Psi_{\rm EP}\}$ is calculated using the correlation between the two ZDC-SMD event planes at forward and backward rapidities~\cite{Poskanzer:1998yz}.
In Au+Au collisions at 27 GeV, the Event Plane Detector (EPD, $|\eta|>3.8$)~\cite{ADAMS2020163970} instead of the ZDC-SMD is used to estimate $\Psi_{\rm{EP}}$, since the latter has a low efficiency. To account for the non-uniform EPD performance, we expand Eq.(\ref{Eq: v1FullEP}) into four terms:
\begin{eqnarray}
v_1^{\rm{f,c}} &=& \frac{1}{\langle \cos^2 \phi \rangle}\cdot  \frac{ \left \langle \cos \phi \cos \Psi^{\rm{f}} \right \rangle}{\sqrt{2\left \langle \cos \Psi^{\rm{f}} \cos \Psi^{\rm{b}} \right \rangle  }},
\label{Eq:vec} \\ 
v_1^{\rm{f,s}} &=& \frac{1}{\langle \sin^2 \phi \rangle}\cdot  \frac{ \left \langle \sin \phi \sin \Psi^{\rm{f}} \right \rangle}{\sqrt{2\left \langle \sin \Psi^{\rm{f}} \sin \Psi^{\rm{b}} \right \rangle  }}, \label{Eq:ves} \\  
v_1^{\rm{b,c}} &=& \frac{1}{\langle \cos^2 \phi \rangle}\cdot  \frac{ \left \langle \cos \phi \cos \Psi^{\rm{b}} \right \rangle}{\sqrt{2\left \langle \cos \Psi^{\rm{f}} \cos \Psi^{\rm{b}} \right \rangle  }}, \label{Eq:vwc} \\
v_1^{\rm{b,s}} &=& \frac{1}{\langle \sin^2 \phi \rangle}\cdot  \frac{ \left \langle \sin \phi \sin \Psi^{\rm{b}} \right \rangle}{\sqrt{2\left \langle \sin \Psi^{\rm{f}} \sin \Psi^{\rm{b}} \right \rangle  }},\label{Eq:vws}
\end{eqnarray}
where $\Psi^{\rm{f}}$ and $\Psi^{\rm{b}}$ are the event planes reconstructed at forward and backward rapidities, respectively. The $\langle \cos^2 \phi \rangle$ and $\langle \sin^2 \phi \rangle$ factors as a function of rapidity compensate for the detector acceptance effect, and they should be constantly $1/2$ for perfect detectors. The event planes from the ZDC-SMD and the EPD are further corrected to have uniform distributions with the method in Ref.~\cite{E877:1997zjw}. The four terms in Eq.(\ref{Eq:vec}-\ref{Eq:vws}) are averaged to give the final $v_1$\{EPD\} results. The event plane reconstructed from spectator nucleons registered by the ZDC-SMD/EPD detectors minimizes the nonflow contributions that are unrelated to the reaction plane orientation and arise from resonances, jets, strings, quantum statistics and final-state interactions.

\renewcommand{\arraystretch}{1.5}
\begin{table}
\centering
\caption{The default and variation cuts in the estimation of  systematic uncertainties for the $v_1$ analyses of Au+Au, Ru+Ru and Zr+Zr events at $\sqrt{s_{\rm NN}}=$ 200 GeV and Au+Au at 27 GeV. }
\label{tab:Systematics}
\begin{tabular}{|c|c|c|} 
\hline
\diagbox[width=80pt]{Systems}{Cuts}                                                              & Default                                                  & Variation                                             \\ 
\hline
\multirow{3}{*}{\begin{tabular}[c]{@{}c@{}}Au+Au \\200 GeV\end{tabular}}            & $-30<V_{z,\rm TPC}< 30$ cm                               & $-30<V_{z,\rm TPC}< 0$ cm                               \\ 
\cline{2-3}
 & $N_{\rm{hits}}\geq 15$                                    & $N_{\rm{hits}}\geq 20$                                   \\ 
\cline{2-3}
  & DCA$\leq 2$ cm                                           & DCA$\leq 1$ cm                                        \\ 
\hline
\multirow{3}{*}{\begin{tabular}[c]{@{}c@{}}Ru+Ru and Zr+Zr \\200 GeV\end{tabular}} & $-35<V_{z,\rm{TPC}}< 25$ cm                               & $-35<V_{z,\rm{TPC}}< 0$ cm                               \\ 
\cline{2-3}
  & $N_{\rm{hits}}\geq 15$                                    & $N_{\rm{hits}}\geq 20$                                   \\ 
\cline{2-3}
  & DCA$\leq 3$ cm                                           & DCA$\leq 2$ cm                                        \\ 
\hline
\multirow{3}{*}{\begin{tabular}[c]{@{}c@{}}Au+Au \\27 GeV\end{tabular}}             & $-70<V_{z,\rm{TPC}}< 70$ cm                              &  $-70<V_{z,\rm{TPC}}< 0$ cm           \\ 
\cline{2-3}
  & $N_{\rm{hits}}\geq 15$                                    & $N_{\rm{hits}}\geq 20$  \\ 
\cline{2-3}
  & DCA$\leq 2$ cm                                           & DCA$\leq 1$ cm      \\
\hline
\end{tabular}
\end{table}

The systematic uncertainties of the $v_1$ measurements are assessed by varying each of the analysis cuts (as listed in Table~\ref{tab:Systematics}) within a reasonable range. 
We estimate the absolute difference ($|\Delta_i|$) between results with the default cut and with a particular cut variation. 
In addition, we also quote the absolute difference  between the $v_1(\mathsf{y})$ slopes measured at forward and backward rapidities as a source of systematic uncertainty. For the analysis of protons, we have examined the effect of protons emitted from the beam pipe in secondary interactions, and find that it contributes less than 1\% to the systematic error. 
The uncertainty due to the particle detection efficiency has been found to be negligible.
The final systematic error is the quadrature sum of the systematic errors from all the  sources under  consideration, each of which is  calculated with $|\Delta_i|/\sqrt{12}$, assuming a uniform probability distribution.
  
\section{Results and discussion}

\begin{figure}[htbp]
\vspace*{-0.1in}
{\includegraphics[scale=0.8]{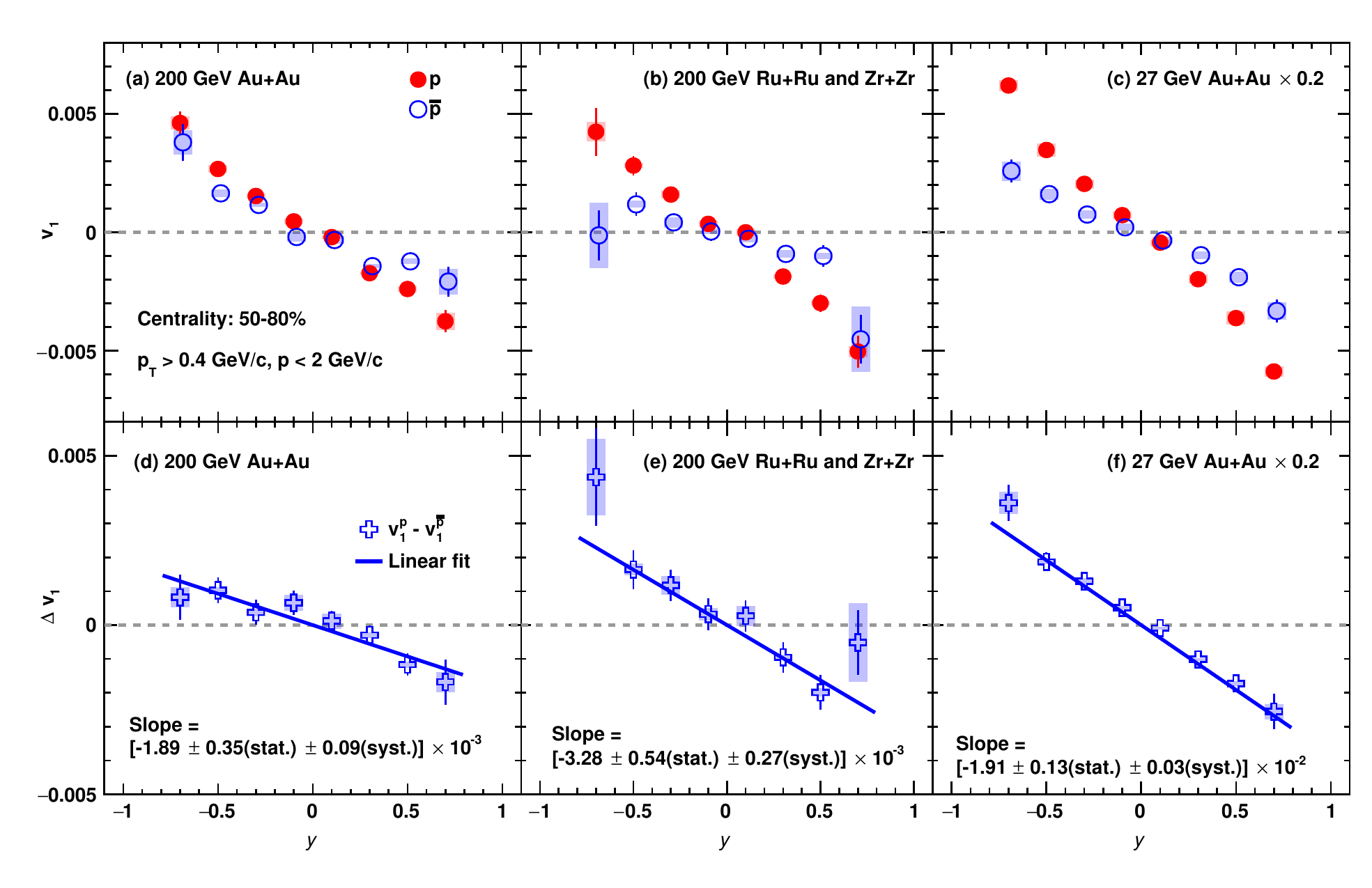}}   	
\captionof{figure}{$v_1$ for protons and anti-protons as a function of rapidity in (a) Au+Au collisions at $\sqrt{s_{\rm NN}}=$ 200 GeV, (b) isobar (Ru+Ru and Zr+Zr) collisions at $\sqrt{s_{\rm NN}}=$ 200 GeV, and (c) Au+Au collisions at $\sqrt{s_{\rm NN}}=$ 27 GeV in the centrality interval of 50--80\%. Protons and anti-protons are marked with solid and open circles, respectively. Panels (d), (e) and (f) show $\Delta v_1 \equiv v_1^{p} - v_1^{\bar{p}}$ versus rapidity. The $d\Delta v_1/d\mathsf{y}$ values are obtained with linear fits (solid lines). Systematic uncertainties are indicated with shaded boxes.}
	\label{fig:v1vsy}
\end{figure}

Figure~\ref{fig:v1vsy} presents $v_1(\mathsf{y})$ for protons and anti-protons in Au+Au collisions at $\sqrt{s_{\rm NN}}=$ 200 GeV, isobar collisions at $\sqrt{s_{\rm NN}}=$ 200 GeV, and Au+Au collisions at $\sqrt{s_{\rm NN}}=$ 27 GeV in the centrality range of 50--80\%. Since the observed difference between the isobaric systems is very small ($\sim 1\sigma$ difference), the Ru+Ru and Zr+Zr data are merged to increase statistics. In general, both protons and anti-protons have negative $dv_1/d\mathsf{y}$ values,
mainly caused by the expansion of  the tilted emission source~\cite{PhysRevC.61.024909}
as demonstrated in Fig.~\ref{fig:illustration2}. In panels (d), (e) and (f) of Fig.~\ref{fig:v1vsy}, we show the $v_1$ difference, $\Delta v_1$, between protons and anti-protons  as a function of rapidity. Linear fits (solid lines) that extrapolate to the origin are applied within $-0.8 < \mathsf{y} < 0.8$, and yield negative $d\Delta v_1/d\mathsf{y}$,  with the significance levels of $5.2 \sigma$ ($5.4 \sigma$) in Au+Au (isobar) collisions at  200 GeV, and $14.3 \sigma$ in Au+Au at 27 GeV. Note that $\Delta (dv_1/d\mathsf{y})$ and $d\Delta v_1/d\mathsf{y}$ are equivalent to each other, and we will stick to the notation, $\Delta(dv_1/d\mathsf{y})$, in the following discussion. This is the first observation of significantly negative $\Delta (dv_1/d\mathsf{y})$ between protons and anti-protons in heavy-ion collisions, and qualitatively agrees with the predictions of the aforementioned electromagnetic field effect, i.e., the dominance of the Faraday induction + Coulomb effect over the Hall and transported-quark effects. The more negative $\Delta (dv_1/d\mathsf{y})$ value at 27 GeV could be partially explained by the slower decay of the spectator-induced electromagnetic field at lower energies due to the longer passage time of incident nuclei. Moreover, the shorter lifetime of the QGP at lower energies causes a stronger remaining magnetic field at the time of chemical freeze-out~\cite{Gursoy:2018yai}. At lower beam energies, anti-protons are more susceptible to annihilation in the baryon-rich environment. Nevertheless, the effect on anti-proton $v_1$ should be marginal since the anti-proton flow measurements~\cite{STAR:2017okv,PhysRevC.93.014907} still meet coalescence expectations.   

\begin{figure}[htbp]
\vspace*{-0.1in}
{\includegraphics[scale=0.8]{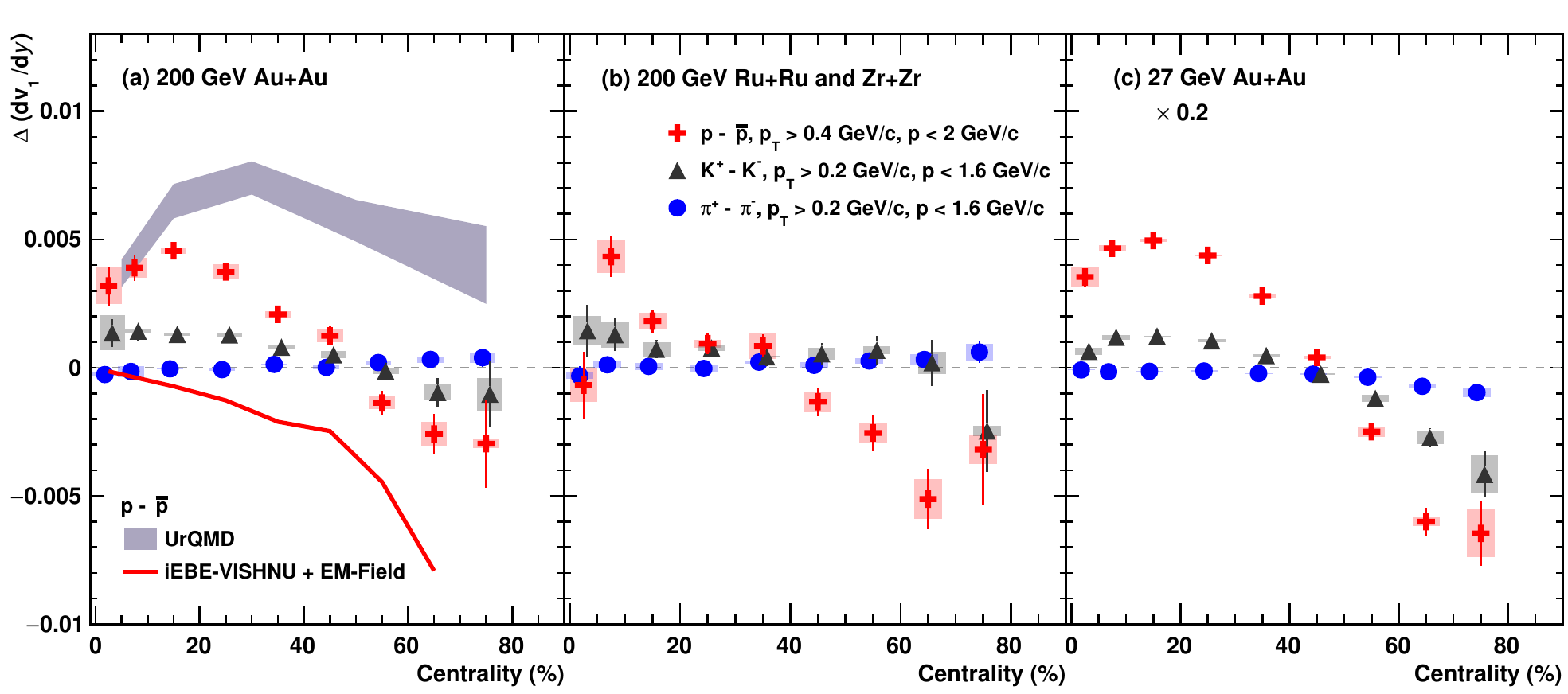}}   	
\captionof{figure}{$\Delta(dv_1/d\mathsf{y})$ between positively and negatively charged pions, kaons and protons as a function of centrality in (a) Au+Au collisions at $\sqrt{s_{\rm NN}}=$ 200 GeV, (b)  isobar collisions at $\sqrt{s_{\rm NN}}=$ 200 GeV and (c) Au+Au collisions at $\sqrt{s_{\rm NN}}=$ 27 GeV.
The lavender band indicates UrQMD simulations, without any EM-field effect, of the proton $\Delta dv_1/dy$ in Au+Au collisions at 200 GeV.
In comparison, a  solid curve is added correspondingly for the iEBE-VISHNU calculation with the electromagnetic field  devoid of transported quarks.~\cite{Gursoy:2018yai}.  }
	\label{fig:SloepVsCent}
\end{figure}

Figure~\ref{fig:SloepVsCent} shows the $\Delta(dv_1/d\mathsf{y}$) between positively and negatively charged hadrons ($\pi^\pm$, $K^\pm$, and $p$($\bar{p}$)) as a function of centrality in Au+Au (a) and isobar (b) collisions at $\sqrt{s_{\rm NN}}=$ 200 GeV, and Au+Au collisions at $\sqrt{s_{\rm NN}}=$ 27 GeV (c). The $\Delta(dv_1/d\mathsf{y})$ values for each particle species are extracted with  the same linear-function fit as in Fig.~\ref{fig:v1vsy}. For all the collision systems and energies, the proton results (closed crosses) display a decreasing trend: positive in central collisions, and  negative in peripheral ones.
This sign change resembles the scenario speculated in Fig.~\ref{fig:Cartoon4Discussion}(c). In central collisions, the magnetic field and the spectator Coulomb field are small since there are few spectator protons, and the dominance of the transported-quark effect leads to the positive $v_1$ splitting.
Towards more peripheral collisions, the electromagnetic field effect becomes stronger,
keeps decreasing $\Delta (dv_1/d\mathsf{y})$, and finally changes the sign. The 
lavender band in Fig.~\ref{fig:SloepVsCent}(a) shows UrQMD simulations~\cite{Bass:1998ca}, which include no electromagnetic fields, and give positive $\Delta(dv_1/d\mathsf{y})$ for protons due to transported-quark contributions.
In Fig.~\ref{fig:SloepVsCent}(a), the solid curve gives the iEBE-VISHNU calculation of the
electromagnetic-field contributions to the proton $\Delta(dv_1/d\mathsf{y})$ without transported quarks in Au+Au collisions at $\sqrt{s_{\rm NN}}=$ 200 GeV~\cite{Gursoy:2018yai}. The iEBE-VISHNU calculation adopts the electrical conductivity of the QGP at equilibrium, $\sigma = 0.023$ fm$^{-1}$, which is estimated from lattice QCD calculations~\cite{Ding:2010ga,Francis:2011bt,brandt2013thermal,Amato:2013naa} with a temperature of $T = 255$ MeV. Since transported quarks are generally believed to provide positive contributions to the proton $\Delta(dv_1/d\mathsf{y})$~\cite{Guo:2012qi,Nayak:2019vtn,Bozek:2022svy}, the negative $\Delta(dv_1/d\mathsf{y})$ values in peripheral events reveal the Faraday induction + Coulomb effect. 
In view of the potential interplay between transported quarks and the electromagnetic field, we abstain from simply adding these two models for a direct quantitative comparison with our data. However, it is noteworthy that the literal sum of the two model outcomes appears to align closely with the measurements, implying that the assumed electrical conductivity lies within a plausible interval.

The decreasing trend and the sign change of $\Delta(dv_1/d\mathsf{y})$ have also been observed between $K^+$ and $K^-$ (closed triangles) in Fig.~\ref{fig:SloepVsCent}, especially at 27 GeV.
Kaons behave in a similar manner as protons, as only $K^+(u\bar{s})$ could be affected by transported $u$ quarks.
Quantitatively, we expect the $\Delta(dv_1/d\mathsf{y})$ for kaons to have a smaller magnitude than that for protons for several reasons. As shown in Fig.~\ref{fig:TPCdedx}(right), kaons have lower mean momentum  and hence lower mean $p_T$ than protons, which can be translated into lower transported quark $v_1$ as well as weaker electromagnetic field effects~\cite{Gursoy:2018yai}.
On average, kaons also have a later formation time than protons due to their lighter mass, and the later-stage quark scatterings could reduce the existing $v_1$ splitting caused by the electromagnetic field. A factor that may complicate the interpretation of the kaon data is the potential asymmetry between $s$ and ${\bar s}$ quarks. For example, the associated strangeness production, $pp \rightarrow p\Lambda(1115)K^+$~\cite{BALEWSKI1998211}, effectively converts net protons (the excess of $p$ over $\bar p$) into $\Lambda(uds)$ and $K^+(u{\bar s})$, and thus $K^+$ receives additional contributions relative to $K^-$. Similar to the charm quarks, the $s(\bar{s})$ quarks are heavier and produced earlier than the $u(\bar{u})$ and $d(\bar{d})$ quarks, and could be dominantly affected by the Hall effect, which reduces the splitting between $K^+$ and $K^-$ in peripheral collisions.

The $v_1$ splitting between $\pi^+$ and $\pi^-$ (closed circles) is less obvious than kaons and protons, but the pion $\Delta(dv_1/d\mathsf{y})$ is  statistically significant,  $-0.0028\pm0.0002$, in 50--80\% Au+Au at 27 GeV. As mentioned before in the discussions related to Fig.~\ref{fig:Cartoon4Discussion},
when transported quarks have positive $dv_1/d\mathsf{y}$, they should give negative contributions to the pion $\Delta(dv_1/d\mathsf{y})$. However, since $\pi^+$ and $\pi^-$ are both affected by transported quarks, the net effect is much smaller than those for kaons and protons due to the cancellation.
In a scenario where transported quarks have negative $dv_1/d\mathsf{y}$~\cite{Parida:2023ldu}, their contribution to 
the $\Delta (dv_1/d\mathsf{y})$ between $\pi^+$ and $\pi^-$ should be positive, and then the negative pion $\Delta (dv_1/d\mathsf{y})$ values at 27 GeV support the dominance of
the Faraday induction + Coulomb effect over the Hall effect. Therefore, the combined $\Delta (dv_1/d\mathsf{y})$ measurements for protons and pions favor the Faraday  + Coulomb effect regardless of the sign of transported-quark $v_1$ in peripheral collisions.
In principle, the electromagnetic effect should give rise to a negative $\Delta(dv_1/d\mathsf{y})$ between $\pi^+$ and $\pi^-$~\cite{Gursoy:2018yai,Zhang:2022lje}, but the aforementioned mechanisms, such as mean $p_T$ and the formation time, are even more severe for pions than for kaons.
The pions from neutral resonance decay may dilute the electromagnetic field effects, whereas the protons from $\Delta^{++}$ decay will enhance this effect as $\Delta^{++}$ has two units of electric charge~\cite{Gursoy:2018yai}. Therefore, the small magnitudes of the pion $\Delta(dv_1/d\mathsf{y})$ are understandable.

\section{Conclusion}
The charge-dependent directed flow provides a probe to the transported quarks, as well as the Hall, Faraday, and Coulomb effects in heavy-ion collisions.
We have presented the $v_1$ measurements for $\pi^\pm$, $K^\pm$, and $p$($\bar{p}$) in Au+Au and isobar (Ru+Ru and Zr+Zr) collisions at $\sqrt{s_{\rm NN}}=$ 200 GeV, and Au+Au collisions at $\sqrt{s_{\rm NN}}=$ 27 GeV. The slope difference, $\Delta(dv_1/d\mathsf{y})$, between protons and anti-protons, as well as between $K^+$ and $K^-$, changes from positive values in central collisions to negative in peripheral collisions. The measured proton $\Delta(dv_1/d\mathsf{y})$ values in the centrality range of 50--80\% are $[-1.89 \pm 0.35({\rm stat.}) \pm 0.09({\rm syst.})] \times 10^{-3}$ in Au+Au collisions at 200 GeV, $[-3.28 \pm 0.53({\rm stat.}) \pm 0.27({\rm syst.})] \times 10^{-3}$ in isobar collisions at 200 GeV, and $[-1.91 \pm 0.13({\rm{stat.}}) \pm 0.03(\rm{syst.})] \times 10^{-2}$ in Au+Au collisions at 27 GeV.
While the positive $\Delta(dv_1/d\mathsf{y})$ for protons and kaons in central collisions can be attributed to the transported-quark contributions, the significant negative values in peripheral events are consistent with the electromagnetic field effects with the dominance of the Faraday induction + Coulomb effect~\cite{Gursoy:2014aka,Gursoy:2018yai,Nakamura:2022ssn}. The observed $v_1$ splitting for protons in Au+Au collisions at $\sqrt{s_{\rm NN}}=$ 200 GeV is comparable in magnitude with the theoretical expectation incorporating both transported quarks~\cite{Guo:2012qi,Nayak:2019vtn,Bozek:2022svy} and the electromagnetic field~\cite{Gursoy:2018yai}. The electrical conductivity of the QGP at equilibrium, $\sigma=$ 0.023 fm$^{-1}$, given by lattice QCD calculations ~\cite{Ding:2010ga,Francis:2011bt,brandt2013thermal,Amato:2013naa} with a temperature of $T=$ 255 MeV, is found to be compatible with the measurements reported in this work.  This charge splitting is stronger in collisions at $\sqrt{s_{\rm NN}}=$ 27 GeV, corroborating the idea that the electromagnetic field decays more slowly at low energies.
Compared with protons, pions and kaons have smaller $\Delta(dv_1/d\mathsf{y})$ magnitudes, which is understandable in view of factors such as mean $p_T$ and the formation time. A companion STAR analysis \cite{STAR:2023wjl}   assumes the coalescence sum rule using combinations of hadrons without transported quarks and concludes that the presence of the EM-field dominated by the Hall effect in mid-central events explains the observed $v_1$ splitting. The combined inference from Ref.~\cite{STAR:2023wjl} and the current work is that a competition between the Hall effect and the Faraday+Coulomb effect, its flavor and centrality dependence may lead to the observed $v_1$ splittings.
Further studies on the beam energy dependence of this observable are underway, with more data accumulated in the RHIC BES-II program.

\section*{Acknowledgement}
We thank the RHIC Operations Group and RCF at BNL, the NERSC Center at LBNL, and the Open Science Grid consortium for providing resources and support.  This work was supported in part by the Office of Nuclear Physics within the U.S. DOE Office of Science, the U.S. National Science Foundation, National Natural Science Foundation of China, Chinese Academy of Science, the Ministry of Science and Technology of China and the Chinese Ministry of Education, the Higher Education Sprout Project by Ministry of Education at NCKU, the National Research Foundation of Korea, Czech Science Foundation and Ministry of Education, Youth and Sports of the Czech Republic, Hungarian National Research, Development and Innovation Office, New National Excellency Programme of the Hungarian Ministry of Human Capacities, Department of Atomic Energy and Department of Science and Technology of the Government of India, the National Science Centre and WUT ID-UB of Poland, the Ministry of Science, Education and Sports of the Republic of Croatia, German Bundesministerium f\"ur Bildung, Wissenschaft, Forschung and Technologie (BMBF), Helmholtz Association, Ministry of Education, Culture, Sports, Science, and Technology (MEXT), Japan Society for the Promotion of Science (JSPS) and Agencia Nacional de Investigaci\'on y Desarrollo (ANID) of Chile.

\textbf{Popular Summary :}

In ultra-relativistic nuclear collisions,  heavy ions such as gold and lead are accelerated close to the speed of light, and pass through each other to create a deconfined  medium of quark-gluon plasma (QGP).
When the collision is non-central, an extremely strong electromagnetic field can be generated in the interaction region.  Although this electromagnetic field has the highest strength  ever achieved on earth, its extremely short lifetime hinders the direct experimental observation.
In this work, we examine the collective motion of charged hadrons emitted out of the medium to uncover the characteristic pattern imprinted by the strong electromagnetic field. We present  unprecedented-precision measurements of charge-dependent sideward motion for $\pi^\pm$, $K^\pm$, and $p$($\bar{p}$) in Au+Au at the nucleon-nucleon center-of-mass energy of 200 GeV and 27 GeV, as well as in isobar (Ru+Ru and Zr+Zr) collisions at 200 GeV. For the first time, a clear signature of the Faraday induction + Coulomb effect has been observed in glancing collisions, which confirms the interaction between 
the strong electromagnetic field and the QGP. 
These results pave the way for investigation of the medium properties and other interesting novel phenomena that are fundamental to our understanding of the strong interaction.

\bibliography{ref}

\end{document}